\documentclass[conference]{IEEEtran}
\usepackage{cite}
\usepackage[pdftex]{graphicx}
\graphicspath{{fig}{images}}
\DeclareGraphicsExtensions{.pdf,.jpeg,.png}
\ifCLASSINFOpdf
  \usepackage[pdftex]{graphicx}
  \graphicspath{{fig}{images}}
  \DeclareGraphicsExtensions{.pdf,.jpeg,.png}
\else
  \usepackage[dvips]{graphicx}
  \graphicspath{{fig}}
  \DeclareGraphicsExtensions{.eps,.png}
\fi
\usepackage[lineno5,leftno]{lgrind}
\begin{document}
\title{Automatic Multi-GPU Code Generation applied to Simulation of Electrical Machines}

\author{\IEEEauthorblockN{A. Wendell O. Rodrigues, Fr\'{e}d\'{e}ric Guyomarc'h\\and Jean-Luc Dekeyser}
\IEEEauthorblockA{LIFL - USTL :: \small{INRIA Lille Nord Europe} - 59650\\Villeneuve d'Ascq - France\\\{wendell.rodrigues,frederic.guyomarch,jean-luc.dekeyser\}@inria.fr}
\and
\IEEEauthorblockN{Yvonnick Le Menach}
\IEEEauthorblockA{L2EP - USTL\\\small{Cit\'{e} Scientifique Bat.P2 - 59655}\\Villeneuve d'Ascq - France\\yvonnick.le-menach@univ-lille1.fr}
}

\maketitle

\begin{abstract}
The electrical and electronic engineering has used parallel programming to solve its large scale complex problems for performance reasons. However, as parallel programming requires a non-trivial distribution of tasks and data, developers find it hard to implement their applications effectively. Thus, in order to reduce design complexity, we propose an approach to generate code for hybrid architectures (e.g. CPU + GPU) using OpenCL, an open standard for parallel programming of heterogeneous systems. This approach is based on Model Driven Engineering (MDE) and the MARTE profile, standard proposed by Object Management Group (OMG). The aim is to provide resources to non-specialists in parallel programming to implement their applications. Moreover, thanks to model reuse capacity, we can add/change functionalities or the target architecture. Consequently, this approach helps industries to achieve their time-to-market constraints and confirms by experimental tests, performance improvements using multi-GPU environments.
\end{abstract}
\IEEEpeerreviewmaketitle

\section{Introduction}
Methods of numerical computing are essential in many scientific and industrial areas. Nevertheless, due to time constraints, communities of those areas are obliged to use parallel platforms to speed-up their results. There are many architectures suitable to parallelize scientific algorithms. Hybrid architectures based on CPU and other devices (e.g. GPU) are popular for economic reasons (i.e. price and energy consumption) and their good performance. However, creating applications on these architectures is an arduous task for non-specialists in parallel programming. This paper presents an approach that addresses:
\begin{enumerate}
	\item \textbf{design methodology} based on MDE to generate automatically application code;
	\item exploiting \textbf{higher performance multi-GPU} validated by a case study.
\end{enumerate}

\section{Background}
A Graphics Processing Unit or GPU is the many-core processor attached to a graphics card. However, though it has diverse cores, its parallelism continues to scale with Moore's law. It is necessary to develop application software that transparently scales its parallelism. Proposals, such as OpenCL, have been designed to overcome this challenge. The Khronos Group released OpenCL\cite{opencl10} as a standard for parallel computing consisting of a language(which is an extension of C), API, libraries and a runtime system. OpenCL is based on a platform model that divides a system into one host and one or several compute devices. Compute devices act as co-processors(e.g. GPUs) to the host(e.g. CPU). An OpenCL application is executed on the host, which sends instructions, defined in special functions called kernels, to the device. Additionally, a single host can have multiple devices. OpenCL allows for creating contexts and queues in order to manage tasks being launched in all attached devices.

\section{Application Design and Code Generation}
\subsection{MDE and MARTE}
Model Driven Engineering (MDE)\cite{lugato10} aims to raise the level of abstraction in program specification and increase automation in program development. The UML profile for MARTE\cite{marte10} extends the possibilities for modeling of application and architecture and their relations. MARTE consists in defining foundations for model-based description of real time and embedded systems.

\subsection{Model Transformation Chain}
In MDE, a model transformation is a compilation process which transforms a source model into a target model. This allows for adding, modifying, transforming model elements in order to achieve a final model closer to the real program application. For instance, the last model has explicit information about variables and task scheduling. In \cite{lugato10} there is an overview about the tools used in \textit{model-to-model} and \textit{model-to-text} process. Additionally, we have used the Gaspard2\cite{gaspard2} framework as the engine to chain and encapsulate these transformations.

\section{Case Study}
\begin{figure*}[!t]
\begin{center}
\includegraphics[scale=.43]{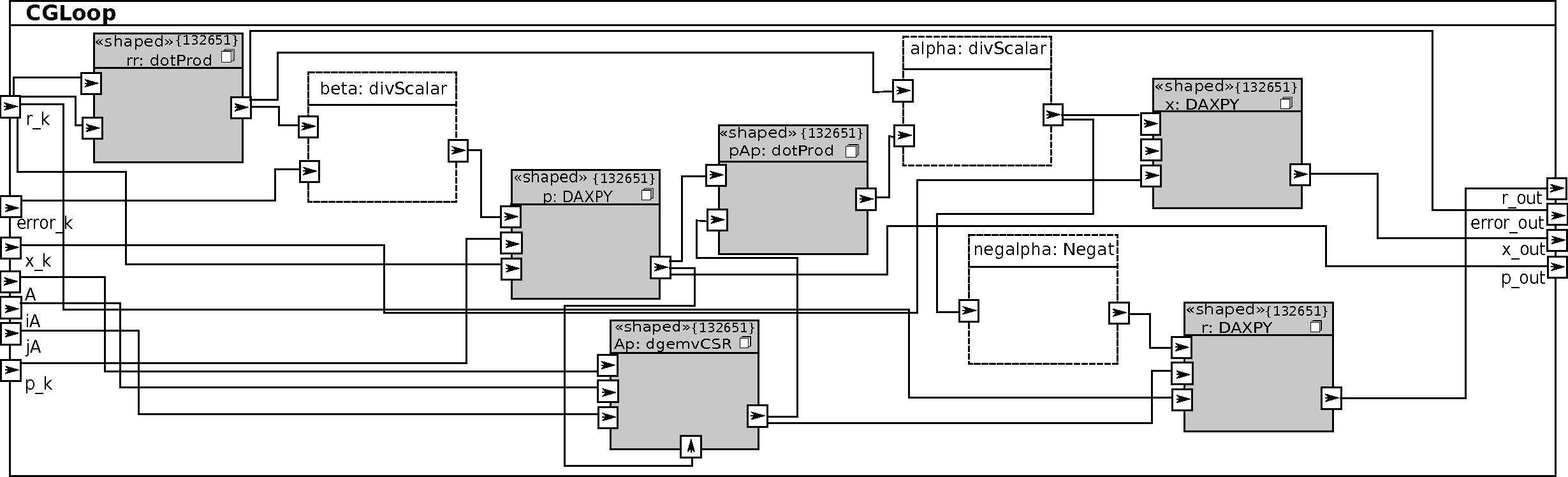}
\end{center}
\caption{Conjugate Gradient UML/MARTE Model}
\label{fig-cgloop}
\end{figure*}

\begin{figure}[!t]
\begin{center}
\includegraphics[scale=.35]{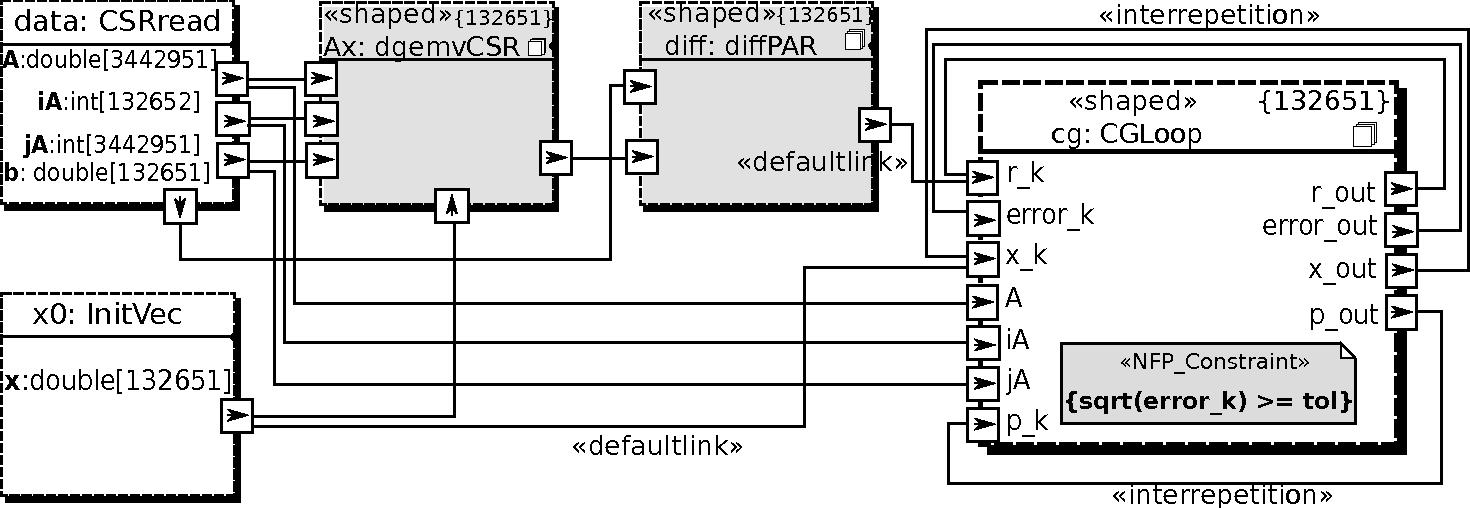}
\end{center}
\caption{ UML/MARTE Model for Setup and CG Overview}
\label{fig-cg}
\end{figure}

The conjugate gradient (CG) method\cite{golub-vanloan:1996} is often used in modeling and simulation of electrical systems. It should only be applied to systems that are symmetric or Hermitian positive definite, and it is still the method of choice for this case. Input data are resulting from a FEM model of an electrical machine. The matrix is stored in \textit{Compressed Sparse Row (CSR)} format having \textit{N}=132651 and \textit{NNZ}=3442951. The CG algorithm is modeled in MARTE as presented in the figure \ref{fig-cg}, where data reading and initial configurations are defined by stereotyped blocks. Highlighted gray blocks represent tasks, which are mapped onto as many devices as we want to distribute the task job. Tasks, such as DGEMV(sparse), are repetitive and, thus, potentially parallel. The CGLoop is a 132651 loop which some of its input data are recovered between continuous iterations. A \textit{continue-condition} is specified by a constraint attached to the CG block, so the loop can stop before running all iterations. The figure \ref{fig-cgloop} is an internal view of the CGLoop modeled in the figure \ref{fig-cg}. Here scalar operations run on CPU processor, and repetitive operations run on GPU processors. Details about deployment of elementary tasks (operations), data and task allocation to architecture, scheduling, grid definition, and so on, can be found in \cite{gaspard2} and they are not discussed in this paper due to scope and space limitation.

\section{Results}
We used four double-precision implementation versions of CG. The first one (the reference) is sequential and uses the Matlab's \textit{pcg} function. The other ones are automatically generated OpenCL implementations whose kernels are launched onto 1, 2 and 4 devices, respectively. The number of used devices depends of the task allocation process. The hardware used is composed by a \textit{2.26GHz Intel Core 2 Duo} processor and S1070 unit (4 Tesla T10 Nvidia GPU). Usually, manually written codes have better performance than automatic ones. However, these automatically generated CG implementations have an expressive performance(table \ref{resultst}) compared to sequential code (time results include just computing and data transfer times in CG loop). The multi-GPU aspect is verified in the two latest versions. The code generation compiler decides equally the task partitioning to the multiple devices. The gain is not linear(though significant) due to extra data transfers among cpu and devices. A detailed analysis about solvers and Multi-GPU can be found in \cite{springerlink:10.1007/s00450-010-0112-6}.

\begin{table}[!ht]
	\centering
	\begin{tabular}{|r|llll|}
                \hline
		conjugate gradient & \#iter & time(s) & speed-up & gflops \\
                \hline
		Matlab PCG & 117 & 3.17 & 1 & .303 \\
		OpenCL (1 GPU) & 116 & 0.659 & 4.81 & 1.45 \\
		OpenCL (2 GPU) & 116 & 0.461 & 6.87 & 2.07 \\
		OpenCL (4 GPU) & 116 & 0.380 & 8.34 & 2.50 \\
                \hline
	\end{tabular}
	\caption{Performance Results; N=132651, NNZ=3442951, tol=1e-10}
	\label{resultst}
\end{table}

\section{Conclusion}
In this paper, we purposed an approach that allows to decrease the application development time for parallel algorithms used in scientific areas. Additionally, the produced code can exploit multi-GPU platforms. Therefore, non-specialists in parallel programming can create applications using the potential power processing of their hybrid architecture. Experimental results show us that this aim is achieved properly for the conjugate gradient method.

\section*{Acknowledgment}
This work is funded by the \textit{Conseil R\'{e}gional Nord-Pas-de-Calais}, Valeo and GPUTech and it's part of the Gaspard2 project, managed by DaRT team of LIFL/INRIA Lille.

\bibliography{compumag2011}
\bibliographystyle{IEEEtran}

\end{document}